# A General Solution for the Implementation of CI/CD in Embedded Linux Development

Behnam Agahi[1] and Hamed Farbeh[2]*

[1]Department of Computer Engineering, Amirkabir University of Technology, Tehran, Iran

[1]behnamagi@tutamail.com, [2]farbeh@aut.ac.ir

*Corresponding Author

## Abstract

With the growing use of embedded systems in various industries, the need for automated platforms for the development and deployment of customized Linux based operating systems has become more important. This research was conducted with the aim of designing and implementing an integrated and reproducible infrastructure for the development, building, and testing of a Linux based operating system using the Yocto Project. The proposed structure was implemented based on a three-layer architecture consisting of the main Yocto repositories, a custom layer (meta custom), and a coordinating manifest layer to ensure version synchronization, scalability, and reproducibility. Three sample projects including libhelloworld, helloworld and the kernel module hello mod were developed and integrated into the build process. Continuous Integration and Continuous Deployment pipelines were implemented with GitLab CI and combined with an isolated Docker environment to automate and streamline the build and testing workflows. Using a local cache server containing hashserv, downloads and sstate cache significantly reduced the build time. The functionality and stability of the system were verified through six boot test scenarios in the QEMU simulator. The results show that the proposed design not only ensures reproducibility but also can be extended to advanced applications such as continuous deployment of real time Linux versions. Future recommendations include expanding automated tests, implementing system monitoring with Prometheus and Grafana, using distributed builds, optimizing with Docker multi stage builds, and enabling continuous deployment of real time Linux changes to provide a stable and scalable model for industrial and research projects in embedded systems with a rapid and reliable development cycle.

# 1. Introduction

With the rapid expansion of digital systems across diverse technological fields, embedded systems have become the computational core of countless intelligent devices and now occupy an essential and undeniable position in modern technology. These systems are integrated combinations of hardware and software designed to perform specific tasks. They are valued for their ability to provide reliable performance, low power consumption, and functionality suited to specialized requirements. Today, they are present in a wide variety of industrial and consumer products, such as telecommunication equipment, smart vehicles, industrial control systems, advanced medical instruments, autonomous robots, intelligent transportation machines, and numerous Internet of Things devices. At the center of most embedded systems operates a lightweight and customized operating system, most often based on Linux, which plays a fundamental role in managing limited resources and executing application software.

A key characteristic of embedded systems is their inherent limitation in processing capacity, memory, power consumption, and real time responsiveness. These factors make software development for such platforms both complex and highly detail sensitive. It requires strict adherence to standardized engineering practices. As applications grow in complexity and the need for continuous updates increases, ensuring optimal performance, reliability, and faster development and delivery cycles has become increasingly important.

Traditional models of embedded software development rely extensively on manual processes and locally configured environments. Such dependence introduces serious difficulties, particularly in large or distributed development teams. Achieving consistent build results, detecting and fixing errors, managing code changes, and performing rapid testing often become slow and error prone tasks. For example, differences in tool versions or environmental configurations between a developer using Windows and another using Linux may cause inconsistent build results or compilation failures. In addition, many testing processes in conventional workflows are conducted manually without uniform standards or systematic documentation. The lack of repeatability in these processes reduces quality and increases maintenance costs.

In response to these challenges, modern development practices known as Continuous Integration and Continuous Deployment have become essential elements of agile software engineering. In this approach, every change in the source code is automatically built, tested, and when successful, delivered or deployed to the target environment. This process accelerates feedback, improves quality control, and enhances collaboration among team members. Although these practices are well established in desktop and web software development, their application in embedded systems continues to face significant technical barriers.

Such barriers include the need to rebuild customized operating systems for each modification, the difficulty of accurate hardware simulation during testing, the complexity of tools such as Yocto, and the limited computing resources available on embedded hardware. Overcoming these obstacles requires an infrastructure that relies on open source and industry grade tools to ensure a fully automated, environment independent, and reliable workflow. The primary goal of this project is to design and implement such an infrastructure for Linux based embedded systems.

This project introduces a complete integrated framework that combines GitLab Continuous Integration pipelines, isolated containers created by Docker, operating system builds managed through Yocto, virtualized testing environments provided by QEMU, and automation scripts implemented in Bash. Together, these components create a unified environment where a developer can initiate the entire build and deployment process simply by submitting a code update, without any dependency on the local configuration. The proposed architecture is modular and easily adaptable to different hardware platforms and software types running on Linux.

Within this framework, Yocto serves as the core component of the operating system build process. It enables customization from the level of kernel configuration and package selection to the structure of the file system through a layered architecture. Developers can define their own meta custom layer to introduce the desired modifications and use Yocto to generate bootable images suited to the specific target architecture. QEMU then executes these images in an emulated environment without requiring any physical hardware, allowing software testing and system validation. The block diagram of the entire system is shown in **Figure 1**.

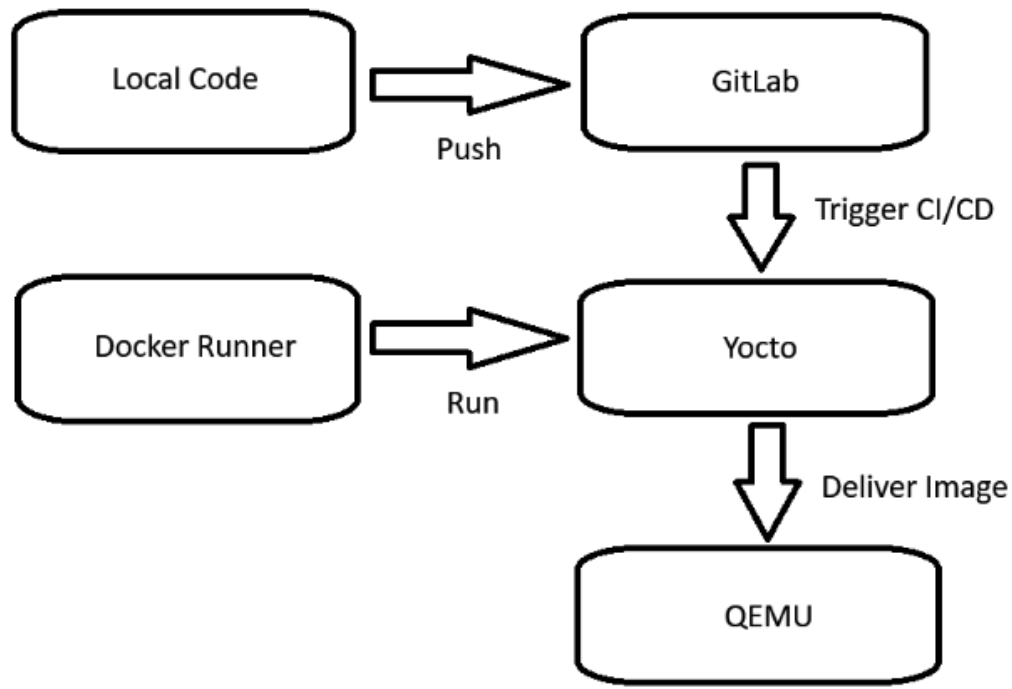

*Figure 1: System's Block Diagram*

The modular structure ensures that each subsystem, including the operating system, applications, kernel modules, and libraries, is defined as an independent project with its own automated pipeline. Any update in one of these components can trigger a complete rebuild at the system level, ensuring consistency and compatibility across all parts of the system. Although the current implementation uses the QEMUARM64 architecture, substituting the appropriate hardware support layer allows adaptation to any other embedded board. The pipeline process is illustrated in **Figure 2**.

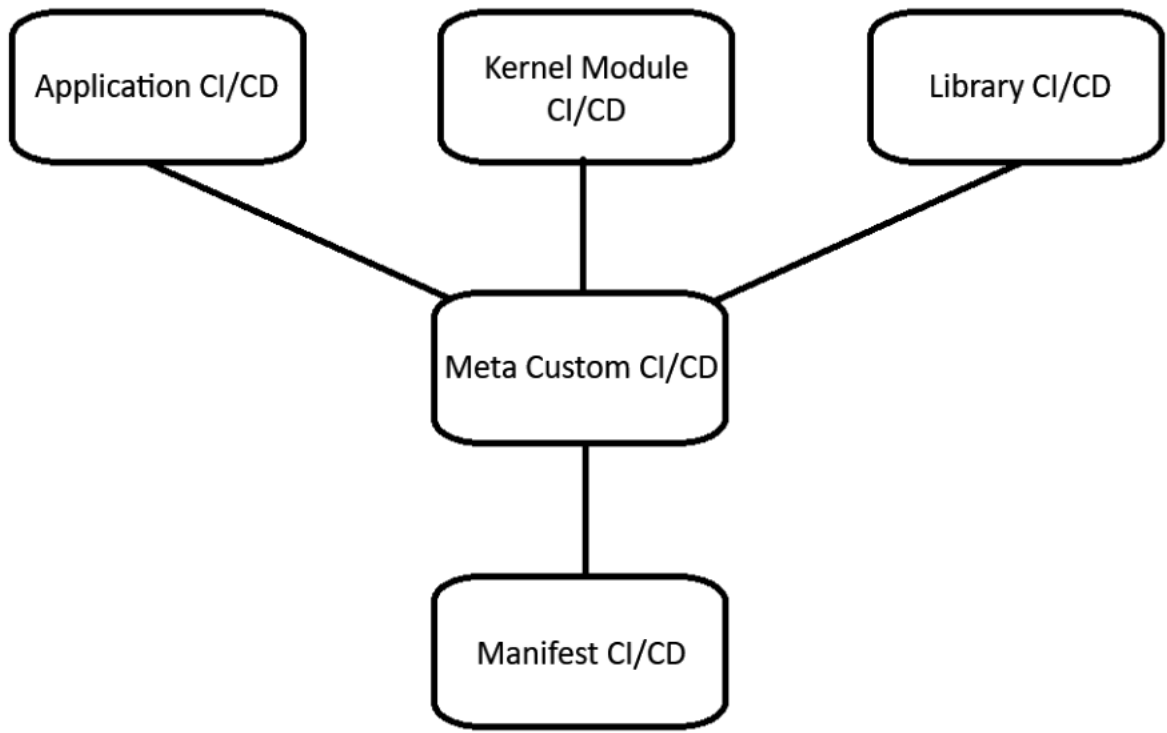

*Figure 2: System's Pipeline Process*

The choice of tools in this project is guided by the need to cover the entire life cycle of Linux based embedded system development according to industrial standards and open source technologies. Yocto provides exceptional flexibility in customizing operating systems, controlling dependencies, and generating optimized images for various processor architectures. Docker creates isolated and reproducible environments, eliminating the impact of local configurations and ensuring consistent build results. GitLab Continuous Integration supplies a unified and standardized platform for executing automated build, testing, and deployment processes while promoting efficient team collaboration. QEMU offers precise hardware emulation, enabling comprehensive testing and debugging without the need for physical prototypes. Bash scripting acts as the connecting element that coordinates and automates all these components. The integration of these tools results in a stable, scalable, and environment independent development pipeline that aligns with professional engineering practices.

In summary, this project represents a practical and forward looking effort to connect modern DevOps methodologies with the specialized domain of embedded system development. The proposed infrastructure provides developers with faster and more reliable processes for development and testing while ensuring environmental consistency and standardization. Through its open source foundation, flexible customization capabilities, and modular design, the system serves as a robust, adaptable, and sustainable solution suitable for both industrial and academic applications.

# 2. Methodology

## *2.1 System Infrastructure Preparation*

Designing and deploying an appropriate infrastructure represents a fundamental and strategic step in any software project, especially in systems whose goal is integration, automation, and continuous deployment of the production process. This infrastructure serves as the foundation upon which all subsequent operations such as development, testing, integration, and deployment are built. Without a reliable, stable, and scalable foundation, even the most accurate code and well defined processes cannot deliver consistent performance. A scientifically designed infrastructure not only maximizes the technical capacity of the development team but also provides a framework that allows the project to evolve and adapt to changing requirements without the need for major redesign.

Based on this principle, the present project adopts a virtualization technology built on ESXi as the central platform. ESXi is a bare metal hypervisor that installs directly on the hardware and eliminates the intermediate host operating system layer, granting complete and direct control over processing, memory, and storage resources. This approach reduces latency, increases performance, and allows precise resource allocation according to the specific needs of each service. The selection of ESXi was motivated by its operational stability, strong support for industrial grade equipment, and compatibility with complex architectures. From a security viewpoint, the advanced virtual networking capabilities of ESXi, including the creation of virtual switches, VLAN configurations, and managed traffic policies, establish clear boundaries between virtual machines and minimize risks of intrusion or data interference. The hardware and software architecture of ESXi is illustrated in **Figure 3**.

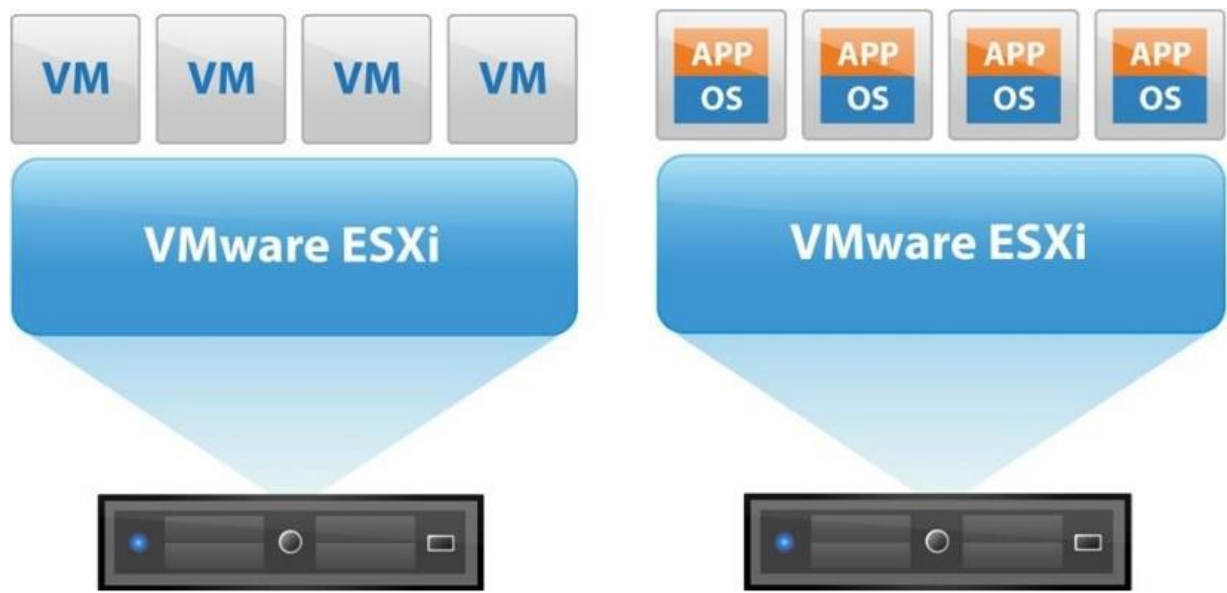

*Figure 3: ESXi Hardware and Software Architecture*

The virtual network structure of the project was designed so that each virtual machine is assigned a fixed and clearly identifiable IP address. This decision not only simplifies management and monitoring but also provides transparency in traffic analysis between services, making troubleshooting faster and more precise when faults occur. A dedicated internal network was established for secure communication between critical services such as GitLab and GitLab Runner, ensuring that the exchange of sensitive data remains fully controlled and isolated from public traffic.

Within this infrastructure, two primary virtual machines were defined, each with a specific and essential responsibility. The first virtual machine hosts GitLab, which serves as the management core and central repository for all source code. This service handles repository management, version control, pipeline execution, and user access management. The second virtual machine hosts GitLab Runner, which acts as the execution agent for the pipelines defined in GitLab and performs the build, run, and test stages of the project. The separation of these two roles not only distributes processing loads efficiently but also improves system security, ensuring that potential failures or attacks on one component do not directly affect the other. The interaction between GitLab and GitLab Runner is shown in **Figure 4**.

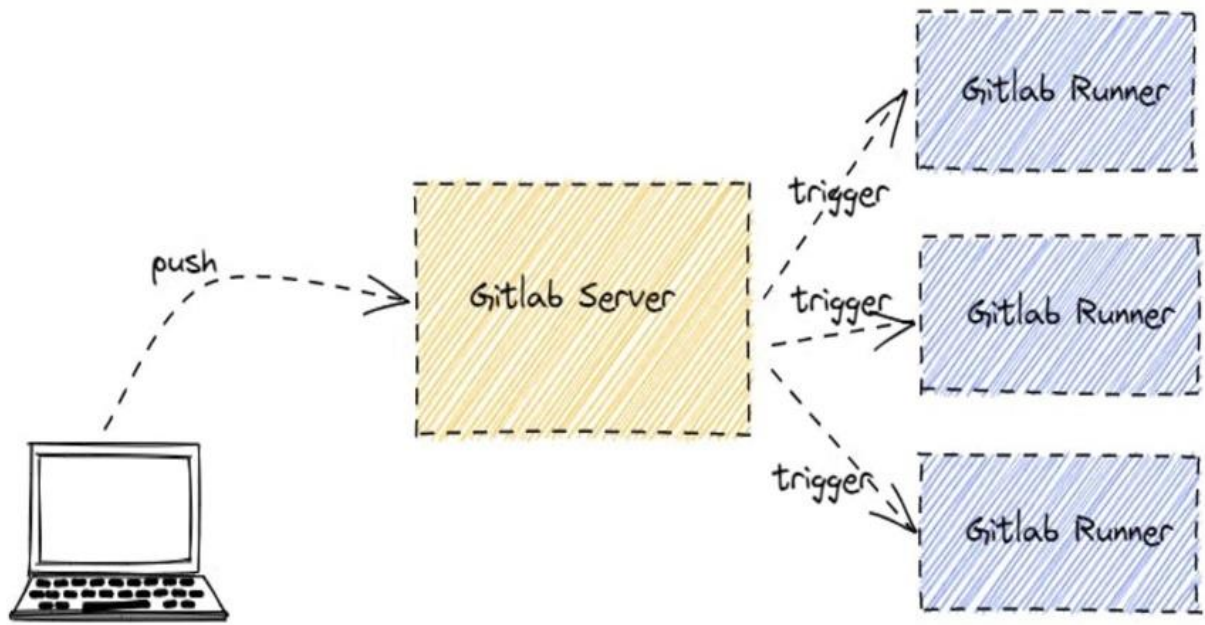

*Figure 4: Gitlab and Gitlab Runner Interaction*

The deployment of GitLab on the first virtual machine was implemented using Docker Compose, which enables the definition and execution of multiple services and dependencies within a single configuration file. This approach guarantees full reproducibility and ensures a consistent operating environment even after restarts or migrations to different servers. Alongside GitLab, the Nginx service was configured as a reverse proxy to handle optimization of communication paths, load management, and data encryption through HTTPS. This configuration not only enhances system security but also improves performance under multiple simultaneous requests.

The second virtual machine runs GitLab Runner, also deployed using Docker Compose. This setup allows the Runner container and its dependencies to operate in an isolated environment without interfering with other services. Registration of the Runner in GitLab was performed using a secure authentication token that confirms the legitimacy of the execution agent. This token is stored safely in configuration files to prevent any information leakage. Once new code changes are pushed to the repository, the Runner automatically initiates the build, test, and deployment processes and reports the results back to GitLab. Nginx again acts as the communication layer, ensuring secure and stable interactions among all components.

Several challenges encountered during this stage highlighted the importance of practical experience and deep technical understanding. For example, synchronizing network configurations between ESXi's virtual network and Docker's internal settings required revising default routes and implementing specific communication rules. Another significant issue involved optimizing the allocation of hardware resources between the two virtual machines since GitLab and Runner can both consume considerable CPU and memory under heavy workloads. Furthermore, version compatibility between GitLab, Runner, and Docker was identified and managed as a critical factor in maintaining system stability.

Aligned with industrial considerations, the infrastructure architecture of this project was designed according to international best practices and standards such as ISO/IEC 27001 for information security and ITIL for IT service management. The modular and isolated design enables each infrastructure component to be updated or replaced without interrupting the overall system operation.

In summary, the infrastructure developed in this project serves not as a passive requirement but as the intelligent backbone of the entire system. By combining ESXi virtualization, Docker based containerization, Docker Compose for service orchestration, and enhanced security through Nginx, a robust foundation is established on which the subsequent chapters of this report will describe the detailed implementation and continuous integration process. This infrastructure satisfies current operational needs while maintaining the capacity to adapt to future requirements and to integrate emerging technologies.

*2.2 Local System Implementation*

The practical phase of this project started with the goal of creating and integrating the software components on the selected platform. The primary framework of this stage was established on the Yocto Project, as it provides a standardized layered structure, simple dependency management through recipes, and the capability to generate customized operating system images.

In the first step, three independent projects were developed so that the key components could be designed, tested, and validated separately. The first project, called helloworld, consisted of a simple program that printed a message on the terminal output. Its main purpose was to verify the performance of the build chain and to ensure the correctness of the initial configuration of the development environment. This project was written in the C programming language, and its build file was configured for full compatibility with the BitBake structure used in the Yocto Project.

The second project, named libhelloworld, was a shared library that implemented the helloworld message as a public function. This library was designed for permanent installation on the Linux operating system and for use by user applications. During installation, the library file was placed in the standard shared library directory of Linux, and the recipe configuration ensured that the helloworld package would link and execute directly from this library. This approach followed the principles of code separation and provided maintainability, code reuse, and the ability to update the library independently.

The third project, named hello mod, was a Linux kernel module integrated into the main build of the operating system by the Yocto Project. In this way, the module was automatically loaded during the boot sequence without requiring any user command. This design aimed to ensure compatibility between kernel level modules and other system components while reducing the complexity associated with manual module loading.

To combine these three projects into the operating system image, a dedicated layer named meta custom was defined within the Yocto Project structure. This layer followed all the standard conventions of the Yocto layer structure and contained the recipes directories for each of the three projects, the conf directory for layer configurations, and the image directory for image customization. Inside each recipe directory, an independent recipe was created, defining the source path, version, dependencies, build process, and installation method.

Within the image configuration section, this layer was set so that the packages produced by the three projects would be added to core image minimal. This capability represents one of the key features of the Yocto Project, enabling customized packages to be appended to a minimal base image and simplifying future system expansion without interference with base components.

For the management of multiple repositories and version synchronization, a separate control project named manifest was designed and implemented. This project included a configuration file named manifest.xml, which contained the name, path, and version of each dependent repository. Using this file, the Google Repo tool fetched all the required repositories simultaneously and placed them in their specified working paths. At this stage, a file named build environment.dockerfile was also prepared to enable the creation of a Docker based build environment in later stages. Although the build process inside the container was not executed at this point and was postponed to the next chapter, preparing this file was a mandatory requirement of the implementation phase.

After fetching the source code and adding the meta custom layer to the Yocto Project structure, the BitBake tool was executed to generate the core image minimal image. Following the defined recipes, this process covered the stages of fetching, configuration, compilation, installation to destination directories, staging, packaging, and final integration into the operating system image. The result was a minimal Linux image containing the helloworld application, the libhelloworld shared library, and the hello mod kernel module preinstalled and automatically loaded by the kernel. The output included the kernel binary and the root file system ready for use. The Yocto build structure is illustrated in **Figure 5**.

```
Sstate summary: Wanted 220 Found 220 Missed 0 Current 932 (100% match, 100% complete)
NOTE: Executing Tasks
NOTE: Tasks Summary: Attempted 3614 tasks of which 3356 didn't need to be rerun and all succeeded.
```

*Figure 5: Yocto Build Structure*

In the functional testing stage, the generated Linux image was executed within QEMU. After successful system boot, running the helloworld program produced the expected message via libhelloworld, confirming that the library was correctly recognized and dynamically linked by the operating system during runtime. In addition, the hello mod kernel module was active from the start of the boot process, and its test message appeared within the kernel log. These results verified that all three components had been properly integrated into the operating system image and were fully functional. The simulation environment and boot process are shown in **Figure 6**.

```
INIT: Entering runlevel: 5
Configuring network interfaces... RTNETLINK answers: File exists
[    8.493859] random: crng init done
[    8.493996] random: 4 urandom warning(s) missed due to ratelimiting
Starting OpenBSD Secure Shell server: sshd
done.
Starting rpcbind daemon...done.
starting statd: done
Starting atd: OK
[   10.425206] Installing knfsd (copyright (C) 1996 okir@monad.swb.de).
starting 8 nfsd kernel threads: [   11.715219] NFSD: Using /var/lib/nfs/v4recovery as the NFSv4 state recovery directory
[   11.716035] NFSD: Using legacy client tracking operations.
[   11.716163] NFSD: starting 90-second grace period (net f0000000)
done
starting mountd: done
Starting system log daemon...0
Starting crond: OK

Poky (Yocto Project Reference Distro) 4.0.13 qemuarm64 ttyAMA0

qemuarm64 login:
```

*Figure 6: Simulation Environment and Boot Process*

In summary, the results of this phase confirm the successful completion of the initial implementation steps, including the creation of the foundational projects, support for a shared library, integration of the kernel module within the system boot process, and successful testing within a virtual simulation environment. The project structure, as built upon the capabilities of the Yocto Project, is thus ready for further enhancement and the addition of new features in subsequent development phases.

*2.3 System Continuous Integration and Continuous Deployment Provision*

After the successful completion of the initial implementation and the generation of the base image using the Yocto Project in the previous chapter, the next phase of the research focused on the full automation of the build and testing cycle. The principal objective of this stage was to establish an infrastructure capable of automatically performing the complete sequence of image generation and verification immediately after any modification in the source code. The process was expected to occur without manual intervention, ensuring accuracy, consistency, and reproducibility in all executions. The chosen platform to meet this requirement was GitLab CI, which offers an effective multi stage pipeline architecture, supports chained dependencies between projects, and enables the automation of build and test processes inside isolated environments. This platform aligned precisely with the modular software and layer based structure of the system.

The implementation process began by defining dedicated pipeline configurations for each of the three individual projects: libhelloworld, helloworld, and hello mod. Within each repository, a configuration file named .gitlab ci.yaml was added. This file described a single pipeline stage responsible for building the project while defining dependency relations to ensure that the output of each lower level project served as input to the next layer. In practice, this meant that changes in the source code or recipe of libhelloworld or hello mod automatically triggered the meta custom pipeline, which then rebuilt its custom layer to include the updated element. This chain based configuration guaranteed that all dependent components remained synchronized and current.

The next essential step was the definition of the pipeline in the meta custom project itself. This project acted as an intermediate stage between the individual components and the global integrator, linking all individual builds together. Its own pipeline, also defined in a .gitlab ci.yaml file, depended directly on the manifest project, which served as the highest level point of coordination and integration. The manifest project combined all layers, recipes, and dependencies to generate the final operating system image. This structured and strictly ordered sequence of dependencies created an automatic propagation chain from the lower modules up to the system integration level. Thus, even a simple code update in the smallest library would automatically trigger the rebuild of the entire Linux image, followed by test execution.

Configuring the pipeline of the manifest project was the most critical part of the system design because it controlled version synchronization and repository management through the Google Repo tool, as well as the actual execution of the BitBake build process. The .gitlab ci.yaml file of this project defined two fundamental stages, executed within a dedicated Docker container environment. The container image was built from the build environment.dockerfile and uploaded to a private GitLab registry. This ensured that every execution of the workflow occurred within an identical and fully reproducible environment, isolating the build from differences in local machine configurations.

In the first stage, two main scripts named repo.sh and image build.sh were executed from the scripts directory of the manifest project. These scripts synchronized all dependent repositories, prepared the build workspace, and invoked BitBake to generate the core image minimal Linux image that contained the three components helloworld, libhelloworld, and hello mod. The configuration guaranteed that the kernel automatically loaded the hello mod module during the startup sequence. After successful completion of the build, the second stage began and ran the image test.sh script, also located in the scripts directory. This script launched the newly generated image inside the QEMU virtualization environment to perform functional testing of the full system.

Because of the high complexity and repetitive nature of the build process, improvement of performance and reduction of build duration were necessary design considerations. To achieve this, a separate virtual machine with a dedicated address was configured to act as a cache server. On this server, a lightweight Python3 service was deployed to provide three key functionalities: bitbake hashserv on port 8001 to maintain build hashes, a downloads repository on port 8002 to store previously fetched sources, and an sstate cache repository on port 8003 to hold reusable intermediate files. The content of these caches, generated after the first full build, was preserved and continuously used in subsequent builds. This reuse significantly decreased both download and compilation times and reduced network traffic and bandwidth usage. The comparison of build durations before and after the cache server deployment is illustrated in **Figure 7**.

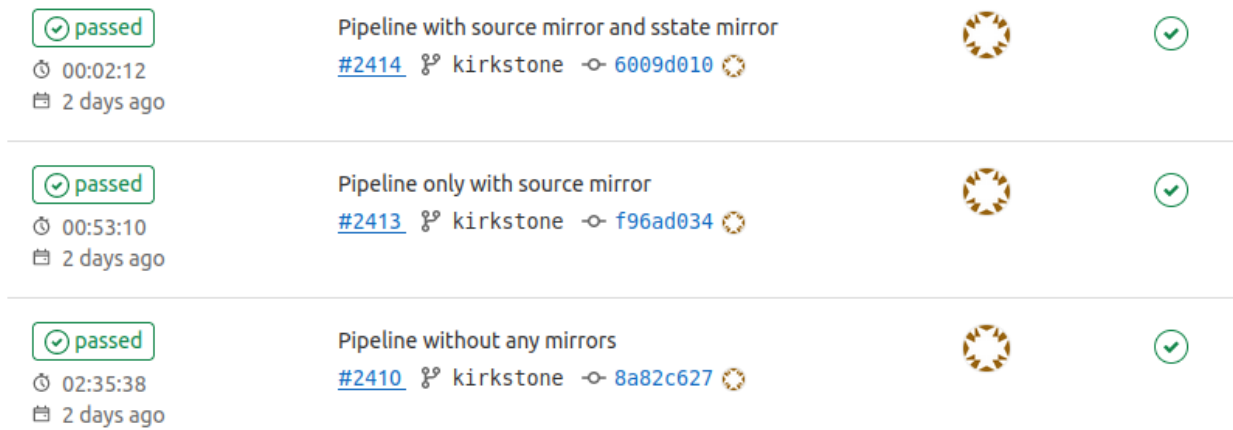

*Figure 7: Build Durations Before and After the Cache Server Deployment*

For formal evaluation and validation of the complete pipeline system, six experimental test scenarios were defined. Each scenario introduced deliberate modifications either in libhelloworld, helloworld, or hello mod to simulate real development activity. These tests included both successful and intentionally failing builds to evaluate error handling and dependency propagation. The observations confirmed that every change correctly propagated through the full hierarchy, triggering dependent pipelines in proper order until the final image was rebuilt and tested automatically. The final successful execution of the complete pipeline sequence is presented in **Figure 8**.

*Figure 8: Final Successful Execution of the Complete Pipeline*

The results of this stage demonstrate that the system achieved a high degree of automation and operational maturity. The integration of GitLab CI pipelines, a containerized build environment, and an independent cache server resulted in substantial improvements in speed, consistency, and reliability. The build and testing process became fully automated, reproducible, and efficient, minimizing manual involvement and human error. This phase thus represents a decisive advancement toward a sustainable and scalable continuous integration and deployment infrastructure fully aligned with the principles and advantages of the Yocto Project.

# 3. Discussion

This research began with a focus on creating an integrated infrastructure for the development, automated building, and testing of Linux systems based on the Yocto Project. Through the previous three chapters, the process advanced from conceptual design and infrastructure preparation to the complete implementation of continuous integration and deployment pipelines and the achievement of functional test results. At the beginning, the main focus was on establishing a stable theoretical and technical framework that could support the requirements of a modular and layered system. This approach, built upon a three-layer architecture consisting of the main Yocto Project layers and their related repositories, the custom meta custom layer, and the coordinating manifest layer, enabled organized communication among all components. Version synchronization and the reproducibility of build outputs were addressed from the very beginning.

Subsequently, practical steps were taken to develop the core components. Three projects were implemented: libhelloworld as a shared library, helloworld as an application linked with that library, and hello mod as a kernel module, all defined inside the meta custom layer to be integrated during the build process. The layer was configured so that the kernel module would be automatically installed and loaded during system startup. The output of this implementation was a Linux image based on core image minimal capable of running in the QEMU emulator and allowing initial functional testing of all components.

After completing the base development phase, the work shifted toward full automation of the build and test cycle. In the third chapter, a sequential and dependent pipeline chain was designed using GitLab CI, with dedicated configuration files for each of the three projects. These configurations ensured that any change in a library, application, or module would automatically trigger the rebuilding of the meta custom layer and finally the complete image of the manifest project. In this project, the build process was divided into two stages, both executed within an isolated and reproducible environment created from build environment.dockerfile. The first stage synchronized repositories through Git Repo and built the image with BitBake, while the second stage executed the image in QEMU.

One of the most significant achievements of this stage was the reduction of build time by employing a separate virtual machine as a cache server that provided bitbake hashserv, downloads, and sstate cache services. This greatly decreased the amount of processing and data transfer required in subsequent builds and improved overall speed. For practical evaluation, six test scenarios were designed covering both successful and failed cases. The results confirmed that the pipelines reacted correctly and that coordination among all layers occurred reliably without error.

In general, the outcome of this research is a system that successfully implements an automated and scalable development and testing cycle within the Yocto Project framework. The system achieves high stability and reproducibility, minimizes build time, and allows continuous and uninterrupted development with complete visibility and control over every stage of the process.

# 4. Further Works

The experience gained from designing and implementing this automated system provides several valuable directions for future improvements and extensions.

The first proposal focuses on enhancing and diversifying functional tests. Although the existing boot tests fully achieved their defined objectives, more advanced evaluations such as software testing, stress testing, security assessment, and compatibility checks with different Linux kernel versions would significantly enhance the quality of validation. Designing scenarios that simulate high computational load, hardware interference, and unstable network conditions could also allow more realistic stability evaluation.

The second proposal involves establishing direct integration between the pipelines and monitoring systems for performance analysis. Connecting pipeline outputs with tools such as Prometheus and Grafana would enable real-time monitoring and analysis of system health, success and failure trends, and build time variations. This capability would be particularly beneficial in projects requiring detailed tracking of release status where rapid identification of issues is essential.

The third direction concerns the use of distributed or parallel building in the Yocto Project to further reduce build time, especially in large projects with many layers or extensive source code. This optimization can be achieved through the use of multiple build hosts, a high-speed central cache server, and effective coordination of compilation tasks. Moreover, adopting Docker multistage builds can minimize the size and dependencies of the final image, producing a more efficient execution environment.

The fourth proposal suggests adding automatic release stages within the pipelines so that once tests are passed successfully, the generated image can be deployed automatically in testing or even production environments. This approach, known in software development as continuous deployment, significantly reduces the time between development and delivery. A more advanced step could include implementing continuous integration and deployment for real-time Linux systems, ensuring that updates to real-time kernel patches and extensions are automatically integrated, built, and tested in the target environments. Such an enhancement would guarantee precision and responsiveness in time-critical applications while extending the reproducibility and orderliness of the release process to the real-time layer itself.

Finally, creating and maintaining comprehensive and up-to-date documentation for the entire system from the methods of layer development and recipe modification to pipeline execution and cache server configuration is of vital importance. These documents facilitate onboarding of new developers, ensure effective knowledge transfer, and prevent errors arising from a lack of technical awareness.

## 5. Conclusion

By implementing the suggested paths, the current infrastructure will not only meet existing project requirements but will also remain capable of expansion toward advanced domains such as real-time Linux systems. The established platform provides a sustainable foundation for automated, traceable, and reproducible embedded Linux development. It stands as evidence that the combination of modular system architecture and continuous integration and deployment methodologies can achieve long-term stability, transparency, and scalability in the development cycle of modern embedded platforms.